\documentclass[twocolumn,aps,pre]{revtex4}
\usepackage{graphicx}
\usepackage{subfigure}
\usepackage{caption2}

\begin{document}

\title{Gap solitons in a model of a hollow optical fiber}
\author{I. M. Merhasin and Boris A. Malomed}
\affiliation{Department of Interdisciplinary Studies, School of Electrical Engineering,
Tel Aviv University, Tel Aviv 69978, Israel}

\begin{abstract}
We introduce a models for two coupled waves propagating in a hollow-core
fiber: a linear dispersionless core mode, and a dispersive nonlinear
quasi-surface one. The linear coupling between them may open a bandgap,
through the mechanism of the avoidance of crossing between dispersion
curves. The third-order dispersion of the quasi-surface mode is necessary
for the existence of the gap. Numerical investigation reveals that the
entire bandgap is filled with solitons, and they all are stable in direct
simulations. The gap-soliton (GS) family is extended to include pulses
moving relative to the given reference frame, up to limit values of the
corresponding boost $\delta $, beyond which the solitons do not exists. The
limit values are nonsymmetric for $\delta >0$ and $\delta <0$. The extended
gap is also entirely filled with the GSs, all of which are stable in
simulations. Recently observed solitons in hollow-core photonic-crystal
fibers may belong to this GS family.
\end{abstract}


\maketitle

The concept of gap solitons (GSs), that was introduced by Mills
\textit{et al}. \cite{Mills}, has drawn a great deal of interest
in optics and other fields where nonlinear waves are important.
After a family of exact solutions for the GSs was found in the
model of a nonlinear optical fiber carrying a Bragg grating (BG)
\cite{exact}, they had been created in the experiment
\cite{experiment}, see further details in reviews \cite{reviews}.
Optical GSs were also predicted in various BG configurations in
the spatial domain \cite{spatial}, and, very recently, solitons of
the gap type have been observed in a discrete array of optical
waveguides \cite{Silberberg} and in Bose-Einstein condensates
\cite{BEC}.

A new optical medium has recently become available in the form of
photonic-crystal fibers (PCFs), with a crystal-like structure of
voids running parallel to the fiber's axis \cite{PCF}. An
important variety of the PCF is a \textit{hollow-core} one, which
provides a unique means for delivery of extremely powerful light
signals, with the energy of up to $100$ pJ, over the distance of
up to $200$ m \cite{Gaeta} - \cite{nearly-zero-dispersion}. The
hollow-core PCF may also be a basic element of soliton-laser
schemes \cite{Wise-laser}. In the general case, theoretical models
for ultra-short solitons in a hollow PCF are complicated, see,
e.g., Ref. \cite{model-equation-two-cycles}. However, in some
cases the experiment \cite{low-loss} and detailed numerical
calculations \cite{Russell,avoided-crossing}\ suggest to use an
approximation which includes a linear dispersionless mode, $v$,
propagating in the gas-filled core, and a quasi-surface one, $u$,
that penetrates into the silica and may therefore be dispersive
and nonlinear. Another medium in which a similar interplay between
the two modes may be expected, is a hollow-core multi-layer fiber;
gap-like (``cutoff") solitons have been recently predicted in it
\cite{Fink}. In this work, we aim to demonstrate that the
coupled-mode system governing the co-propagation of the core and
quasi-surface waves may open a spectral bandgap, which gives rise
to a two-parameter family of stable GSs. The family is a novel
one, being quite different from the well-studied standard example
\cite{reviews}.

A straightforward derivation, following the well-known principles of the
coupled-mode theory \cite{reviews,Agr}, makes it possible to cast coupled
equations for local amplitudes of the above-mentioned dispersive and
dispersionless modes in the following normalized form (details of the
derivation will be presented elsewhere):
\begin{eqnarray}
iu_{z}-icu_{\tau }+\frac{1}{2}u_{\tau \tau }+i\gamma u_{\tau \tau \tau
}+|u|^{2}u+v &=&0,  \label{u} \\
iv_{z}+icv_{\tau }+u &=&0.  \label{v}
\end{eqnarray}Here, $z$ and $\tau $ are, as usual, the propagation distance and reduced
time for the unidirectional propagation, the effective coefficient
of the Kerr nonlinearity in the $u$-mode and inter-mode linear
coupling constant are scaled to be $1$, the second-order
dispersion of the $u$-mode is assumed anomalous, and its
coefficient is also normalized to be $1$, and $2c$ is the
group-velocity mismatch between the two modes [the respective
coefficients in Eqs. (\ref{u}) and (\ref{v}) are made symmetric
($\mp ic$) by adjusting the reference frame; in fact, a
group-velocity asymmetry will be re-introduced below, when looking
for moving solitons]. Note that phase-velocity terms can be
eliminated, and the coefficients in front of $iu_{z}$ and $iv_{z}$
can be made equal by means of obvious linear transformations,
which is implied in the equations.

Because the solitons observed in the hollow-core PCFs are very
short, with the temporal width $\sim 200$ fs, and the carrier
wavelength may be close to the zero-dispersion point
\cite{nearly-zero-dispersion}, the model must take into regard the
third-order dispersion \cite{Gaeta} -
\cite{nearly-zero-dispersion}, which is accounted for by the
coefficient $\gamma $ in Eq. (\ref{u}). Thus, the normalized
system contains two free parameters, $c$ and $\gamma $.

Without displaying details of the derivation, we note that $\Delta
z=1$ and $\Delta \tau =1$ in the normalized units correspond,
typically, to $10$ cm and $100$ fs in the real-world PCFs, and the
nonlinear coefficient in Eq. (\ref{u}), set equal to $1$, may
actually range between $10^{-6}$ and $10^{-8} $ W$^{-1}$cm$^{-1}$
in physical units \cite{Russell,Gaeta}.

An effect that the model does not include is the soliton's self-frequency
shift due to the intra-pulse stimulated Raman scattering (generally, it is
an essential effect for ultra-short solitons \cite{Agr}). From the
experiment, it is known that this effect may be both significant and
negligible in the hollow-core PCFs, depending, in particular, on the choice
of the core-filling gas \cite{Gaeta}. In this Letter, we focus on the
simplest conservative model which does not include the self-frequency shift;
its role will be considered in detail elsewhere. We do not include either
the fourth-order dispersion, self-steepening, intrinsic nonlinearity of the
core mode, and loss. Estimates demonstrate that these factors are not
expected to change the results dramatically; in particular, the net loss
accumulated over the experimentally relevant transmission length is $\sim 1$
dB \cite{Russell,Gaeta}.

First, it is necessary to find the spectrum of the linearized
version of the system. Substituting $\left( u,v\right) \sim \exp
\left( ikx-i\omega \tau \right) $, we obtain a dispersion relation
that can be easily solved, yielding two branches,\begin{eqnarray}
4k_{1,2} &=&-\omega ^{2}(1+2\gamma \omega )\pm   \label{spectrum} \\
&&\sqrt{\omega ^{4}(1+2\gamma \omega )^{2}+8c\omega ^{2}(2c+\omega
+2\gamma \omega ^{2})+16}.  \nonumber
\end{eqnarray}Further consideration demonstrates that, once we set $c>0$ by definition,
the spectrum (\ref{spectrum}) gives rise to a bandgap, where
solitons may exist, only in the case of $\gamma >0$, which we
assume below (this case is quite possible, as PCFs allow one to
engineer the dispersion of the guided modes). The formation of the
bandgap is actually stipulated by the principle of the avoidance
of crossing between dispersion curves of the decoupled equations
(\ref{u}) and (\ref{v}); it is known that this principle is a
crucial factor in the explanation of dispersion properties of the
PCFs \cite{Russell,avoided-crossing}.

\begin{figure}[tbp]
\centering\includegraphics[width=3in]{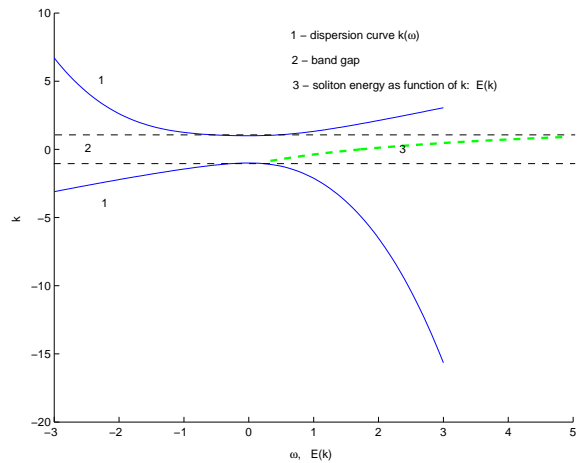}
\caption{The solid
curves are two branches of the dispersion relation
(\protect\ref{spectrum}), for $c=1$, $\protect\gamma =0.3$. The
bandgap is located between the branches. The bold dashed curve
inside the bandgap shows the dependence $E(k)$ for the soliton
family found in the gap. As is seen, the family fills the entire
bandgap, and it always satisfies the VK stability condition,
$dE/dk>0$. }\label{Fig1}
\end{figure}

A typical example of the spectrum that contains the bandgap is
shown in Fig. \ref{Fig1}. The gap-existence region covers nearly
the entire quadrant $\left( c>0,\gamma >0\right) $ in the system's
parameter plane. We also note that the model never gives rise to
more than one gap. An asymptotic expansion of Eq. (\ref{spectrum})
shows that, for small $c$, the minimum value of the
third-order-dispersion coefficient, which is necessary for the
existence of the bandgap, scales so that $\gamma _{\min }\sim
c^{1/3}$.

In the gap, stationary soliton solution were sought for by the
substitution of $\left\{ u(z,\tau ),v(z,\tau )\right\} =\left\{
U(\tau ;k),V(\tau ;k)\right\} \exp \left( ikz\right) $ in Eqs.
(\ref{u}) and (\ref{v}) and solving the resulting ODEs by means of
the relaxation numerical method. Note that the functions $U$ and
$V$ in this ansatz cannot be assumed real; however, they obey
symmetry restrictions, so that $\mathrm{Re}\left( U(\tau )\right)
$ and $\mathrm{Re}\left( V(\tau )\right) $ are even functions of
$\tau $, while $\mathrm{Im}\left( U(\tau )\right) $ and
$\mathrm{Im}\left( V(\tau )\right) $ are odd. Stability of the GS
solutions was then tested in direct simulations. Additionally, it
was tested by means of the Vakhitov-Kolokolov (VK) criterion
\cite{VK}: for the soliton family, the condition for the stability
against perturbations with real eigenvalues is $dE/dk>0$, where
the energy is
\begin{equation}
E(k)=\int_{-\infty }^{+\infty }\left[ \left\vert U(\tau ;k)\right\vert
^{2}+\left\vert V(\tau ;k)\right\vert ^{2}\right] d\tau .  \label{E}
\end{equation}

\begin{figure}[tbp]
\centering\subfigure[]{
\includegraphics[width=3in]{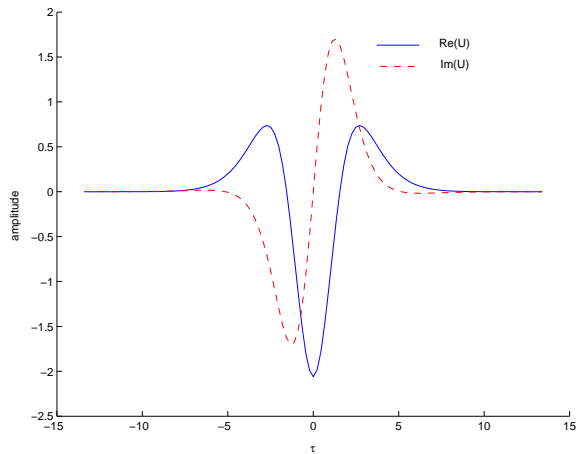}}\newline
\centering\subfigure[]{
\includegraphics[width=3in]{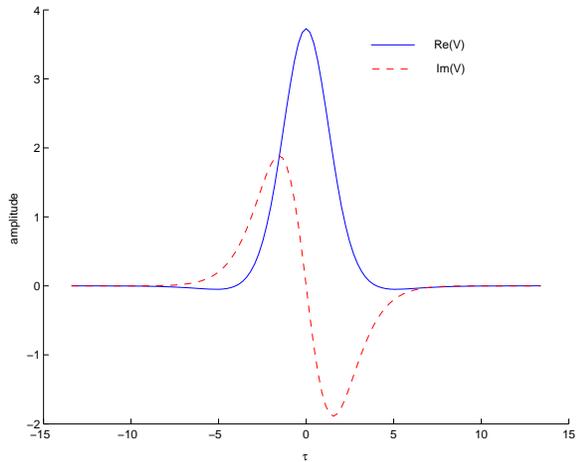}}
\caption{A generic example of the gap soliton (with $k=0$ $\ $and
$\protect\delta =0$) found inside the bandgap shown in Fig.
\protect\ref{Fig1}. Note that the absolute values of the fields,
$\left\vert U(\protect\tau )\right\vert $ and $\left\vert
V(\protect\tau )\right\vert $, always have an ordinary
single-humped shape. } \label{Fig2}
\end{figure}

\begin{figure}[tbp]
\centering\includegraphics[width=3.5in]{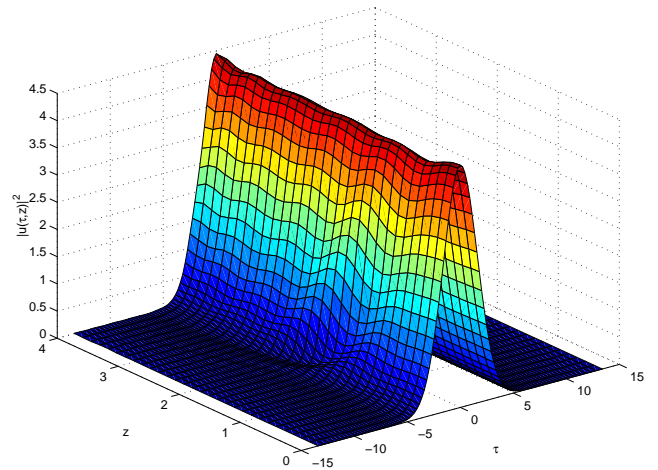} \caption{Stable
evolution of the soliton from Fig. \protect\ref{Fig2}, to which an
arbitrary initial perturbation was added, whose amplitude is $2\%$
of the soliton's amplitude. The evolution is of the field
$\left\vert v(z,\protect\tau )\right\vert $ is similar to that
displayed here for $\left\vert u(z,\protect\tau )\right\vert $.}
\label{Fig3}
\end{figure}

As a result, it has been concluded that the entire bandgap is
filled by the solitons, and they all are stable in direct
simulations (extensive scanning of the parameter space has not
turned up any example of an unstable soliton). Besides that, all
the GSs satisfy the VK criterion, see Fig. \ref{Fig1}. A typical
shape of the stationary GS is displayed in Fig. \ref{Fig2}, and
its stability against finite perturbations is illustrated by Fig.
\ref{Fig3}.


A natural extension of the GS family is generated by lending the
soliton a finite velocity, in the reference frame of Eqs.
(\ref{u}) and (\ref{v}). Note that, as well as the ordinary GS
model, the present one features no Galilean (or other) invariance
that would generate moving solitons automatically. To develop this
extension, we rewrite Eqs. (\ref{u}) and (\ref{v}) in the boosted
reference frame, replacing the independent variables $\left(
z,\tau \right) $ by $\left( z,\tau ^{\prime }\equiv \tau -\delta
~z\right) $, where $\delta $ is the velocity shift. The
transformation replaces the terms $i\left( u_{z},v_{z}\right) $ by
$i\left( u_{z}-\delta ~u_{\tau },v_{z}-\delta ~v_{\tau }\right) $
and, accordingly, in the dispersion relation (\ref{spectrum}) $k$
is replaced by $k+\delta ~\omega $. As a result, the bandgap
changes its shape with the increase of $\delta $.

\begin{figure}[tbp]
\centering\includegraphics[width=3in]{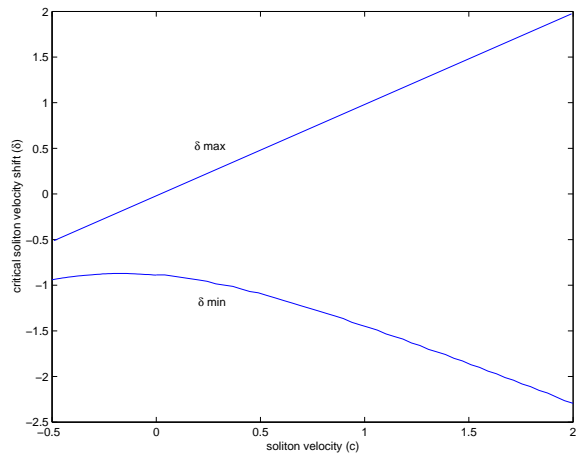} \caption{In the
reference frame boosted by the velocity shift $\protect\delta $,
the bandgap exists in the interval $\protect\delta _{\min
}<\protect\delta <\protect\delta _{\max }\equiv c$. The two lines
show $\protect\delta _{\min }$ and $\protect\delta _{\max }$ vs.
$c$.} \label{Fig4}
\end{figure}

Naturally, the gap closes when $|\delta |$ is too large, and it is
necessary to find the corresponding maximum and minimum values of
the velocity shift, $\delta _{\max }$ and $\delta _{\min }$, up to
which the bandgap exists. Straightforward analysis of the
dispersion relation demonstrates that $\delta _{\max }=c$, which
has a simple explanation: the above transformation replaces the
coefficient $c$ in Eq. (\ref{v}) by $c-\delta $, and the gap
closes down when this combination vanishes. The other limit value
can be found numerically from the corresponding algebra. The final
result is presented in Fig. \ref{Fig4}, that shows $\delta _{\max
}$ and $\delta _{\min }$ as functions of $c$ for given $\gamma $.
A noteworthy feature of this diagram is that the gap-supporting
region continues, for $\delta \neq 0$, to negative $c$.

Solving the stationary equations in the boosted reference frame numerically,
we have concluded that the entire gap is again completely filled with
solitons. The boosted solitons obey the same symmetry constraints as in the
case of $\delta =0$, so that the real and imaginary parts of the stationary
fields are, respectively, even and odd functions of $\tau ^{\prime }$. The
fields $\left\vert u(z,\tau ^{\prime })\right\vert $ and $\left\vert
v(z,\tau ^{\prime })\right\vert $ are always single-humped. A characteristic
example of the boosted soliton in shown in Fig. 5. Finally, adding arbitrary
initial perturbations shows that all the moving solitons are stable, in
direct simulations, everywhere in the bandgap.
\begin{figure}[tbp]
\centering\subfigure[]{\includegraphics[width=3in]{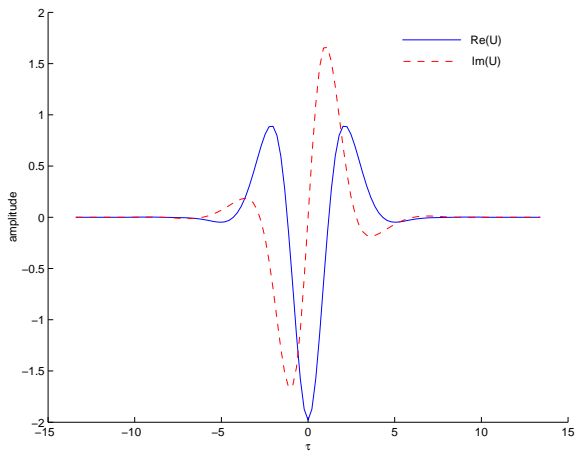}}\newline
\centering\subfigure[]{\includegraphics[width=3in]{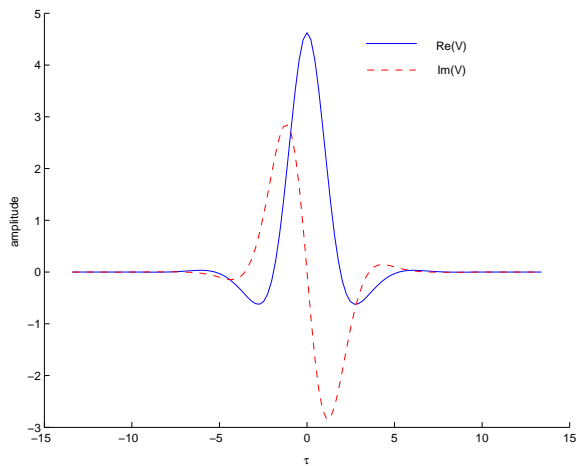}}
\caption{A generic example of the boosted gap soliton with
$\protect\delta =-0.5$, found at $c=0$ and $\protect\gamma =0.3$.
Note that, for $c=0$, the solitons with $\protect\delta =0$ do not
exist, see Fig. \protect\ref{Fig4}. In terms of $\left\vert
u\right\vert $ and $\left\vert v\right\vert $, the boosted
solitons always have a single-humped shape.}
\end{figure}

In conclusion, we have proposed the simplest model for two co-propagating
waves in the hollow-core fiber, one a linear dispersionless core mode, and
the other a dispersive nonlinear quasi-surface mode. The linear coupling
between them may open a bandgap, through the mechanism of the crossing
avoidance. The third-order dispersion of the latter mode is a necessary
condition for the existence of the gap. The numerical results demonstrate
that the entire bandgap is filled with solitons, and they all are stable in
direct simulations. The gap-soliton (GS) family has been extended to include
the boosted solutions, up to the limit values of the boost $\delta $ (these
values are different for positive and negative $\delta $).

Solitons which were recently observed in hollow-core
photonic-crystal fibers may belong to this GS family. The
theoretical results suggest to extend the experimental study of
these solitons, and of the ``cutoff" solitons in hollow-core
multilayer fibers, in order to identify their plausibly GS nature.
On the other hand, the theoretical model needs to be extended by
inclusion of the self-frequency shift and some other features.
Results of the extension will be reported elsewhere.

After the submission of this paper for publication, a work by D.V.
Skryabin, Opt. Exp. 12, 4841 (2004) has appeared, which considers
a very similar model and also reports finding stable solitons in
it.

\end{document}